\documentclass[conference]{IEEEtran}

\usepackage{comment}
\usepackage{fancyvrb}
\usepackage{graphicx}
\usepackage[ragged]{footmisc}
\usepackage{ragged2e}
\usepackage[section]{placeins}
\usepackage{hyperref}

\begin{document}

\title{In the transmission of information, the great potential of model-based coding with the SP theory of intelligence}

\author{\IEEEauthorblockN{J Gerard Wolff}
\IEEEauthorblockA{CognitionResearch.org\\
Menai Bridge\\
UK\\
Email: jgw@cognitionresearch.org\\
Telephone: +44 1248 712962}}

\maketitle

\begin{abstract}

Model-based coding, described by John Pierce in 1961, has great potential to reduce the volume of information that needs to be transmitted in moving big data, without loss of information, from one place to another, or in lossless communications via the internet. Compared with ordinary compression methods, this potential advantage of model-based coding in the transmission of data arises from the fact that both the transmitter (``Alice'') and the receiver (``Bob'') are equipped with a grammar for the kind of data that is to be transmitted, which means that, to achieve lossless transmission of a body of data from Alice and Bob, a relatively small amount of information needs to be sent. Preliminary trials indicate that, with model-based coding, the volume of information to be sent from Alice to Bob to achieve lossless transmission of a given body of data may be less than $6\%$ of the volume of information that needs to be sent when ordinary compression methods are used.

Until recently, it has not been feasible to convert John Pierce's vision into something that may be applied in practice. Now, with the development of the {\em SP theory of intelligence} and its realisation in the {\em SP computer model}, there is clear potential to realise the three main functions that will be needed: unsupervised learning of a grammar for the kind of data that is to be transmitted using a relatively powerful computer that is independent of Alice and Bob; the encoding by Alice of any one example of such data in terms of the grammar; and, with the grammar, decoding of the encoding by Bob to retrieve the given example. It appears now to be feasible,  within reasonable timescales, to bring these capabilities to a level where they may be applied to the transmission of realistically large bodies of data.

\end{abstract}




\IEEEpeerreviewmaketitle

\section{Introduction}


``The Square Kilometre Array is one of the most ambitious scientific projects ever undertaken. Its organizers plan on setting up a massive radio telescope made up of more than half a million antennas spread out across vast swaths of Australia and South Africa.'' So say John Kelly and Steve Hamm, both of IBM, in their book {\em Smart Machines} \cite[p.~62]{kelly_hamm_2013}.

Their reason for writing about the SKA is that it will create huge problems for even the smartest or most powerful of smart machines. ``The SKA is the ultimate big data challenge.'' say Kelly and Hamm. ``The telescope will collect a veritable deluge of radio signals from outer space---amounting to fourteen exabytes of digital data per day …'' ({\em ibid}., p.~63). Of the several problems arising from quantities of data like that, one that may seem surprising is that the amount of energy required merely to move the data from one place to another is proving to be a significant headache for the SKA project and other projects of that kind.

More generally: ``Communication networks face a potentially disastrous `capacity crunch'\thinspace''\footnote{This quote summarises the conclusions of a meeting organised by the UK's Royal Society, introduced in \cite{ellis_et_al_2015}, with other papers from the meeting at \href{http://bit.ly/2fSy6qN}{bit.ly/2fSy6qN}.} and ``Internet access may soon need to be rationed because the UK power grid and communications network can not cope with the demand from consumers.''\footnote{From a website of the telecomms company BT (\href{http://bit.ly/2eUfMbS}{bit.ly/2eUfMbS}).}

These problems may be solved or at least reduced via a new approach to old ideas: ``analysis/synthesis'' and, more specifically, the relatively challenging idea of ``model-based coding''. This paper expands and develops the relatively brief discussion in \cite[Section VIII]{sp_big_data}.

Analysis/synthesis has been described by Khalid Sayood like this:

\begin{quote}

    ``Consider an image transmission system that works like this. At the transmitter, we have a person who examines the image to be transmitted and comes up with a description of the image. At the receiver, we have another person who then proceeds to create that image. For example, suppose the image we wish to transmit is a picture of a field of sunflowers. Instead of trying to send the picture, we simply send the words `field of sunflowers'. The person at the receiver paints a picture of a field of sunflowers on a piece of paper and gives it to the user. Thus an image of an object is transmitted from the transmitter to the receiver in a highly compressed form.'' \cite[p.~592]{sayood_2012}.

\end{quote}

This approach works best with the transmission of speech, probably because the physical structure and properties of the vocal cords, tongue, teeth, and so on, help in the process of creating an analysis of any given sample of speech and in any synthesis of speech that may be derived from that analysis. But things are more difficult with images, especially if they are moving.

The concept of model-based coding was described by John Pierce in 1961 like this:

\begin{quote}
    ``Imagine that we had at the receiver a sort of rubbery model of a human face. Or we might have a description of such a model stored in the memory of a huge electronic computer. First, the transmitter would have to look at the face to be transmitted and ‘make up’ the model at the receiver in shape and tint. The transmitter would also have to note the sources of light and reproduce these in intensity and direction at the receiver. Then, as the person before the transmitter talked, the transmitter would have to follow the movements of his eyes, lips and jaws, and other muscular movements and transmit these so that the model at the receiver could do likewise.'' \cite[pp.~139--140]{pierce_1961}.

\end{quote}

At the time this was written, it would have been impossibly difficult to make things work as described. Pierce says: ``Such a scheme might be very effective, and it could become an important invention if anyone could specify a useful way of carrying out the operations I have described. Alas, how much easier it is to say what one would like to do (whether it be making such an invention, composing Beethoven’s tenth symphony, or painting a masterpiece on an assigned subject) than it is to do it.'' ({\em ibid}., p.~140).

Even today, Piece's vision is a major challenge. But there appears to be a way forward, described in the rest of this paper. With some development, it is likely to be very effective in the lossless transmission of big data and in lossless communications via the internet.

\begin{figure*}[!htbp]
\centering
\includegraphics[width=0.9\textwidth]{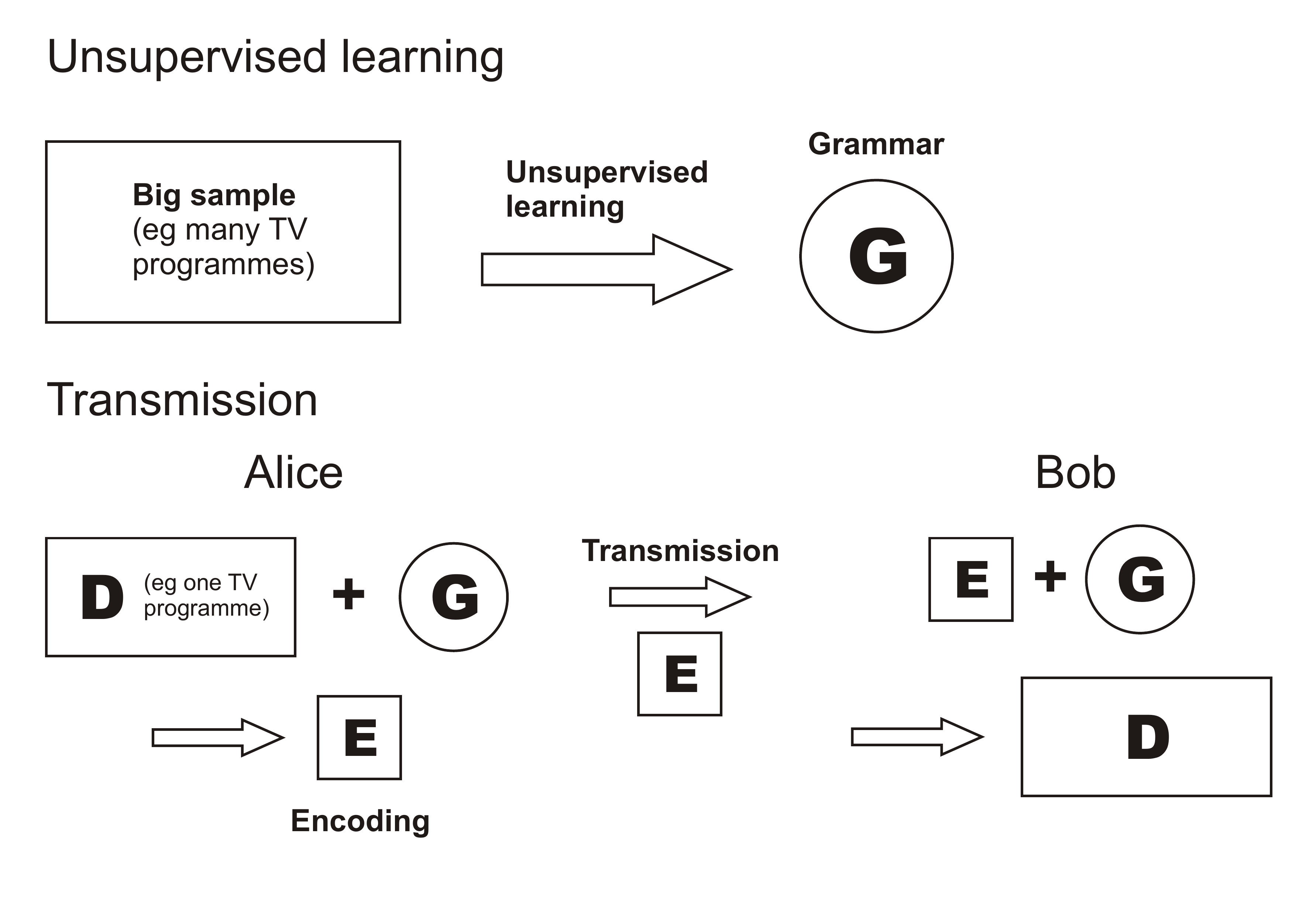}
\caption{A schematic view of how, with model-based coding, information may be transmitted efficiently from Alice to Bob.}
\label{learning_transmission_figure}
\end{figure*}

In outline, model-based coding may be made to work as shown in Figure \ref{learning_transmission_figure}. There would be two main elements to the scheme: learning of an abstract description or `grammar' (`{\bf G}') for the kind of information to be transmitted; and transmission of information from A (`Alice') to B (`Bob').

The learning would be ``unsupervised'', meaning learning directly from data without assistance of any kind of ``teacher'', or the labelling of examples, or rewards or punishments, or anything equivalent. Learning would normally be done independently of any specific transmission, it would be done by a relatively powerful computer, and with a relatively large sample of the kind of data that is to be transmitted, such as a large collection of TV programmes.

Alice and Bob would each receive a copy of {\bf G}. For example, {\bf G} may be installed on every new computer and every new smartphone, and it may also be made available for downloading.

In transmission of any one body of information (`{\bf D}'), such as one TV programme, {\bf D} would first be processed by Alice in conjunction with {\bf G} to create an `encoding' (`{\bf E}') which would describe {\bf D} in terms of the entities and abstract concepts in {\bf G}. The encoding, {\bf E}, would then be transmitted to Bob who would use it, in conjunction with his own copy of {\bf G}, to reconstruct {\bf D}. Provided that Alice and Bob have the same {\bf G}, the version of {\bf D} that is created by Bob should be exactly the same as the version of {\bf D} that was transmitted by Alice, without loss of information.

Since {\bf E} would normally be very small compared with {\bf D}, there would normally, with one qualification, be a large saving in the amount of information to be transmitted compared with the transmission of raw data. Also, for reasons given below, it is likely that {\bf E} would normally, and with the same qualification, be very small compared with what would be transmitted using ordinary compression methods such as LZ, JPEG or MPEG, without the benefit of model-based coding.

The qualification is that any given {\bf G} would be used for the transmission of many different {\bf D}s. If {\bf G} is only used once or twice, any saving is likely to be relatively small because it would be reduced by the cost of transmitting {\bf G} to Alice and Bob---unless {\bf G} is pre-installed in new computers and smartphones as suggested above when the marginal cost of distribution would be small.

The main differences between model-based coding and alternative schemes using ordinary compression methods are these:

\begin{enumerate}

    \item Any ``learning'' with ordinary compression methods is part of the encoding stage, not an independent process. The result of compressing any one body of data {\bf D} may be seen to comprise a grammar derived from {\bf D} which we shall call {\bf Gx}, together with an encoding of {\bf D} in terms of {\bf Gx} which we shall call {\bf Ex}---although those two elements may be not clearly separated.

    \item Any such learning with ordinary compression methods is normally relatively unsophisticated and designed to favour speed of processing on low-powered computers rather than high levels of information compression.

    \item With ordinary compression methods, Alice transmits both {\bf Gx} and {\bf Ex} together, not {\bf E} by itself. As we shall see, this is likely to mean much smaller savings than if {\bf E} is transmitted alone.

    \item In some versions of MPEG compression, Alice and Bob may be provided with some elements of {\bf G}---such as the structure of human faces or bodies---but these are normally hard coded and not learned.

        Any learning in this case appears to be within a framework that lacks generality, is restricted to such things as faces or bodies, and is without the potential for unsupervised learning of a wide variety of entities and concepts (see, for example \cite{feng_etal_1999,moghaddam_pentland_1995,aizawa_etal_1989}).

        It appears that after the year 2000, few if any researchers have been conducting research on model-based coding, perhaps because of the difficulties that John Pierce anticipated.

\end{enumerate}

Of these four points, the third is the most important and provides the key to relatively large gains in efficiency in transmission with model-based coding compared with transmission with ordinary methods for information compression.

To develop transmission of information via model-based coding as outlined above, a promising way forward is via the {\em SP theory of intelligence}, outlined in the \href{sp_overview_appendix}{Appendix}. This system, the product of a long-term programme of research, has clear potential to provide the main functions that are needed: unsupervised learning of {\bf G}; encoding of {\bf D} in terms of {\bf G}; and lossless recreation of {\bf D} from {\bf E} and {\bf G} \cite[Section VIII]{sp_big_data}.

If the SP system is being used by Alice as a means of transmitting information economically to Bob, then, with a previously-learned {\bf G} playing the part of Old knowledge and a given body of information ({\bf D}) playing the part of New information, the encoding created by the system may play the part of {\bf E} in the transmission of {\bf D}, as described above.

Regarding the first of the functions mentioned above---unsupervised learning of {\bf G}---the SP computer model has already demonstrated unsupervised learning of plausible generative grammars for the syntax of English-like artificial languages, including the learning of segmental structures, classes of structure, and abstract patterns \cite[Chapter 9]{wolff_2006}. With non-linguistic or ``semantic'' forms of knowledge, the system has clear potential to learn such things as class hierarchies, class heterarchies (meaning class hierarchies with cross classification), part-whole hierarchies, and other forms of knowledge \cite[Section 9.5]{wolff_2006}.

A key idea in the SP framework is that the entities and abstract concepts discovered by the system would be ``natural'' in the sense that they would be the kinds of things that people recognise, including specific things like ``my cat'' and more general concepts like ``animal''. Evidence to date suggests that the SP system conforms to this principle---{\em the discovery of natural structures via information compression}, or ``DONSVIC'' for short \cite[Section 5.2]{sp_extended_overview}. It appears that unsupervised learning in accordance with the DONSVIC principle yields relatively high levels of information compression.

Preliminary results with the SP computer model show that the size of {\bf E} can be less than $6\%$ ($0.56$) of the total size of {\bf Gx} and {\bf Ex} together. In other words, {\em transmission of information with model-based coding is likely to be {\em \bf very} much more efficient than transmission of information using ordinary compression methods}.

When the SP system has been generalised to process 2D patterns, it is anticipated that unsupervised learning in the SP system may be extended to the learning of 3D digital models of objects, in much the same way that some existing applications can build such models, each one from overlapping digital photographs of an object taken from different angles \cite[Section 6.1]{sp_vision}. The SP system should also be able to build 3D digital models of environments from overlapping images, much as Google Streetview builds what are essentially 3D models of streets, using overlapping photographs \cite[Section 6.2]{sp_vision}.

The second of the functions mentioned above---encoding of {\bf D} in terms of {\bf G}---is accommodated in the way the system builds multiple alignments from New information (received from the system's environment) and Old knowledge (that is derived via earlier learning and is stored for current and future use). As noted in the \href{sp_overview_appendix}{Appendix}, a key part of that process is the creation of a relatively compact encoding of the New information in terms of the Old knowledge.

Regarding the third of the functions mentioned above---recreation of {\bf D} from {\bf E} and {\bf G}---a neat feature of the SP system is that decoding of information is done in exactly the same way as the encoding of information \cite[Section 3.8]{wolff_2006}, with {\bf E} playing the part of New information and, as before, {\bf G} playing the part of Old knowledge. So it is a straightforward matter for Bob to use the SP system to decode any {\bf E} received from Alice, using his own copy of {\bf G}.

\section{Conclusion}

Model-based coding, described by John Pierce in 1961, has great potential to reduce the volumes of data that need to be transmitted in moving big data from one place to another or in communications via the internet.

Instead of transmitting a `grammar' and an `encoding' of the data to be transmitted in terms of the grammar---which, with minor deviations, is what is needed with ordinary compression methods---it is only necessary to transmit an encoding of the data. This advantage of model-based coding arises from the fact that, by contrast with the use of ordinary compression methods in the transmission of data, both Alice and Bob are equipped with a grammar for the kind of data that is to be transmitted.

Preliminary trials indicate that the volume of information to be transmitted with model-based coding may be less than $6\%$ of the volume of information to be transmitted with ordinary compression methods.

Until recently, it has not been feasible to convert John Pierce's vision into something that may be applied in practice. Now, with the development of the SP system, there is clear potential to realise the three main functions that will be needed: unsupervised learning of a grammar for the kind of data that is to be transmitted; the encoding of any one example of such data in terms of the grammar; and decoding of the encoding to retrieve the given example.

It appears now to be feasible to develop these capabilities within reasonable timescales. By contrast with other work on model-based coding, unsupervised learning in the SP system has the potential to learn what will normally be the great diversity of entities and concepts that are implicit in the data.

With these developments, big data may glide quickly and efficiently from one place to another, without the need for massive bandwidth, and without needing the output of a small power station to haul it on its way. And there may be less need to worry about possible shortages of bandwidth in the internet or shortages of energy to power the internet.

\appendix

\section{The SP theory and the SP computer model}\label{sp_overview_appendix}

In outline, the SP theory, and its realisation in the SP computer model, has been designed to simplify and integrate observations and concepts across artificial intelligence, mainstream computing, mathematics, and human perception and cognition \cite{sp_extended_overview,wolff_2006}. The SP system has distinctive features and advantages compared with other AI-related systems, including deep learning in neural networks \cite{sp_alternatives}.

The system comprises these main features:

\begin{enumerate}

    \item All kinds of knowledge are represented with arrays of atomic symbols in one or two dimensions called {\em patterns}. At present, the SP computer model works only with one-dimensional patterns but it is envisaged that it will be generalised to work with patterns in two dimensions. Two-dimensional patterns may serve in the representation of 3D structures \cite[Section 6.1 and 6.2]{sp_vision}.

    \item All kinds of processing are done via a process of searching for patterns or parts of patterns that match each other and via the merging or ``unification'' of patterns, or parts of patterns, that are the same---with a consequent compression of information.

    \item More specifically, all kinds of processing are done via the building and manipulation of {\em multiple alignments}, a concept borrowed and adapted from bioinformatics. An example of a multiple alignment from bioinformatics is shown in Figure \ref{dna_multiple_alignment_figure}.

    \item The whole system is inherently probabilistic because of the very close connection that is known to exist between information compression and concepts of prediction and probability \cite{li_vitanyi_2014}.

\end{enumerate}

\begin{figure*}[!hbt]
\fontsize{10.00pt}{12.00pt}
\centering
{\bf
\begin{BVerbatim}
  G G A     G     C A G G G A G G A     T G     G   G G A
  | | |     |     | | | | | | | | |     | |     |   | | |
  G G | G   G C C C A G G G A G G A     | G G C G   G G A
  | | |     | | | | | | | | | | | |     | |     |   | | |
A | G A C T G C C C A G G G | G G | G C T G     G A | G A
  | | |           | | | | | | | | |   |   |     |   | | |
  G G A A         | A G G G A G G A   | A G     G   G G A
  | |   |         | | | | | | | |     |   |     |   | | |
  G G C A         C A G G G A G G     C   G     G   G G A
\end{BVerbatim}
}
\caption{An example of a multiple alignment of the kind created in bioinformatics, with 5 short sequences of DNA bases.}
\label{dna_multiple_alignment_figure}
\end{figure*}

The main difference between multiple alignment in bioinformatics and multiple alignment in the SP system is that, in the first case, all sequences have the same status, whereas in the SP system, one of the patterns (sometimes more than one) is designated as {\em New}, while the other patterns are designated {\em Old}, and the system is designed to search for one or more multiple alignments that will yield a relatively economical encoding of the New pattern or patterns in terms of one or more of the Old patterns. This encoding takes the form of an SP pattern and its size in bits is input to a measure of success in the compression of the New pattern.

The concept of multiple alignment, as it has been developed in the SP programme of research, has proved to be very versatile and powerful. It may provide the long-sought-after key to general AI, meaning AI with the versatility and adaptability of human intelligence. I believe it is fair to say that it could be the ``double helix'' of intelligence---as significant for an understanding of ``intelligence'' in a broad sense as is DNA for the biological sciences.

In keeping with the quest for simplification and integration across a broad canvass, the SP system has strengths in several different areas including: unsupervised learning, the representation and processing of diverse kinds of knowledge; the processing of natural language, fuzzy pattern recognition, recognition at multiple levels of abstraction, best-match and semantic forms of information retrieval, several kinds of reasoning, planning, and problem solving.

\sloppy The SP system also has several potential benefits and applications described in peer-reviewed papers that may be downloaded via links from \href{http://www.cognitionresearch.org/sp.htm}{www.cognitionresearch.org/sp.htm}.

It is envisaged that an {\em SP machine}, derived from the SP computer model, will be developed as a high-parallel software virtual machine, hosted on an existing high-performance computer. This will be a means for researchers everywhere to see what can be done with the SP system and create new versions of it.

\bibliographystyle{IEEEtran}

\end{document}